# Highly effective relativistic free carrier plasma mirror confined within a silicon slow light photonic crystal waveguide

Mahmoud A. Gaafar,[1,2,*] Dirk Jalas,[1] Liam O'Faolain,[3,4,5] Juntao Li,[6,+] Thomas F. Krauss,[7] Alexander Yu. Petrov[1,8], and Manfred Eich[1,9]

[1]*Institute of Optical and Electronic Materials, Hamburg University of Technology, Hamburg 21073, Germany*
[2]*Department of Physics, Faculty of Science, Menoufia University, Menoufia, Egypt*
[3]*SUPA, School of Physics and Astronomy, University of St. Andrews, St. Andrews, Fife KY16 9SS, United Kingdom*
[4]*Tyndall National Institute, Lee Maltings Complex, Dyke Parade, Cork, Ireland*
[5]*Centre for Advanced Photonics and Process Analysis, Cork Institute of Technology, Cork, Ireland*
[6]*State Key Laboratory of Optoelectronic Materials & Technology, Sun Yat-sen University, Guangzhou 510275, China*
[7]*Department of Physics, University of York, York, YO105DD, United Kingdom.*
[8]*ITMO University, 49 Kronverkskii Ave., 197101, St. Petersburg, Russia*
[9]*Institute of Materials Research, Helmholtz-Zentrum Geesthacht, Max-Planck-Strasse 1, Geesthacht, D-21502, Germany*
*\*Corresponding author: mahmoud.gaafar@tuhh.de,*
*+Corresponding author for slow light sample fabrication: lijt3@mail.sysu.edu.cn*

**Abstract**

Reflection at relativistically moving plasma mirrors is a well-known approach for frequency conversion as an alternative to nonlinear techniques. A key issue with plasma mirrors is the need for a high carrier concentration, of order $10^{21}$ cm$^{-3}$, to achieve an appreciable reflectivity. To generate such high carrier concentrations, short laser pulses with extreme power densities of the order of $\geq 10^{15}$ W/cm² are required. Here, we introduce a novel waveguide-based method for generating relativistically moving plasma mirrors that requires much lower pump powers and much less carrier concentration. Specifically, we achieve an experimental demonstration of 35% reflection for a carrier concentration of $5 \cdot 10^{17}$/cm³ generated by a power density of only $1.2 \cdot 10^9$ W/cm². Both the plasma mirror and the signal are confined and propagating within a solid state silicon slow light photonic crystal waveguide. This extraordinary effect only becomes possible because we exploit an indirect intraband optical transition in a dispersion engineered slow light waveguide, where the incident light cannot couple to other states beyond the moving front and has to reflect from it.

The moving free carrier (FC) plasma mirror is generated by two photon absorption of 6 ps long pump pulse with a peak power of 6.2 W. The reflection was demonstrated by the interaction of a continuous wave (CW) probe wave co-propagating with the relativistic FC plasma mirror inside a 400 µm long slow light waveguide. Upon interaction with the FC plasma mirror, the probe wave packets, which initially propagate slower than the plasma mirror, are bounced and accelerated, finally escaping from the front in forward direction. The forward reflection of the probe wave packets are accompanied by a frequency upshift. The reflection efficiency is estimated for the part of the CW probe interacting with the pump pulse.

Our novel concept bears various technical advantages such as that no vacuum conditions and no ultrahigh power lasers are required. Most importantly, it allows efficient reflection at small FCs concentration. These advantages, in addition to the fact that the described effects take place using guided waves "on-chip", will allow an exploitation of the relativistic plasma mirror effect in integrated optics technology. Furthermore, due to the fact that the pump, probe and shifted probe are all at 1.5 µm wavelength, the presented effect opens new possibilities for frequency manipulation and all optical switching in optical telecommunication.

# Introduction

Reflection of electromagnetic waves from moving relativistic mirrors has been of interest for many years due to its potential for efficient frequency conversion[1]. However, relativistic plasma mirrors require a high concentration of free carriers (FCs) to achieve an appreciable reflectivity for optical frequencies. For the plasma to act as a metallic reflector, the plasma density $N$ must be high enough for the effective plasma frequency $\omega_p$ to exceed the frequency of the input wave $\omega_i$, where $N = \omega_p^2 \varepsilon_0 \varepsilon_r m / e^2$ [2,3]. For a wavelength of 800 nm for example, this corresponds to a carrier density of $1.7 \cdot 10^{21}$ cm$^{-3}$, while 1550 nm requires $0.46 \cdot 10^{21}$ cm$^{-3}$. Moving mirrors can be realized as electron bunches flying with relativistic velocities. Different schemes have been developed to create such relativistic structures [4–7]. The highest concentration so far, of the order of $10^{21}$ cm$^{-3}$, has been demonstrated in electron bunches generated by ionizing targets using electromagnetic ponderomotive forces under blow-out conditions[5–7]. These require short laser pulses with extreme power densities of the order of $10^{18}$ W/cm². However, due to relativistic effects, the FC density is smaller in the reference frame moving together with the electron bunch and therefore orders of magnitudes larger FC densities as $10^{21}$ cm$^{-3}$ are required to provide complete reflection [6].

Ionization fronts are a more versatile route to achieving efficient reflection. In this case, only the boundary of a free carrier plasma is moving. This approach is well known for frequency conversion in free space at microwave frequencies[2,8–11]. More recently, the effect has been extended to THz light by further increasing the FC concentration and generating the plasma front in a silicon wafer [3,12].

We now translate this exciting physics into guided wave optics and show how the waveguide dispersion can be exploited to achieve a substantial increase in reflection. Photonic crystal (PhC) waveguides, fabricated on silicon, are an appealing platform due to their compatibility with on-chip integration and their inherent flexibility in dispersion design. First, the ability to dramatically enhance nonlinear effects[13–15] by using slow light PhC waveguides[16–19] facilitates the generation of FCs by two photon absorption (TPA) using picojoule pulses. Furthermore, slow light PhC waveguides allow for tailoring the photonic bands with regions of different group velocities and dispersion[20,21]. This way, an ionization front can be generated by a pump moving with a group velocity that is different to the group velocity of the signal [22,23]. The transmission of light through the ionization front can then be described as an indirect photonic transition, whereby the wave vector and frequency of an optical probe are simultaneously altered upon a transition between two modes of a photonic structure[22,24,25]. Recently, the reflection from a counter propagating front in a slow light waveguide was theoretically proposed assuming a FC concentration of the order of $10^{19}$/cm³, which is not accessible by TPA in silicon[26]. Here, we optimize the indirect transition for the copropagating scheme and adjust the group velocity of the pump that generates FC plasma. We manage to achieve the case when light stays in the same band of the unaltered waveguide and thus undergoes an indirect *intraband* transition. As a result, we demonstrate a reflectivity of 35% from the plasma mirror. The moving plasma mirror is generated by the two photon absorption of a 6 ps long pump pulse with a peak power of 6.2 W; this peak power corresponds to an intensity of $1.2 \cdot 10^9$ W/cm² and an induced FC density of $5 \cdot 10^{17}$/cm³. We demonstrate the high reflectivity via the interaction of a continuous wave (CW) probe wave co-propagating with the plasma mirror inside a 400 µm long slow light waveguide. Upon interaction with the plasma mirror, the probes, which we consider to be composed of wave packets, initially propagate slower than the plasma mirror, are bounced and accelerated, finally escaping from the front in forward direction. The forward reflection of the probe wave packets (i.e. the "bouncing") is accompanied by a frequency upshift. In this respect, the effect is similar to the optical analogue of event horizons induced by Kerr nonlinearity[27–29]. We also show that the FC plasma mirror is asymmetric in its reflection properties. It will accelerate and reflect a slow probe wave packet propagating in front of it, yet it will transmit a fast probe wave packet approaching from behind.

**Approach**

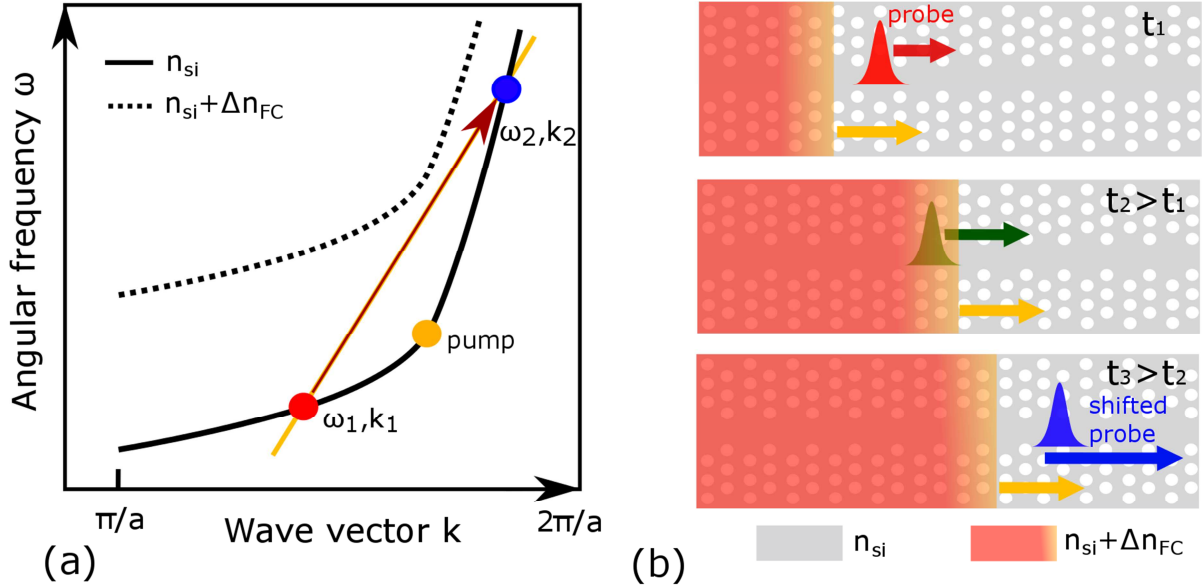

Figure 1: (a) Schematic representation of an intraband indirect photonic transition in a slow light silicon PhC waveguide. The solid curve represents the dispersion band of waveguide mode in the ground state (with refractive index $n_{si}$), while the dashed curve indicates the switched state with refractive index $n_{si} + \Delta n_{FC}$. We use the photonic crystal band with positive group velocity to represent the dispersion relationship. The orange line represents the phase continuity line with a slope equal to the group velocity of the pump pulse set at the knee of the dispersion band (orange dot). The red and blue dots indicate the initial (slow mode) and final (fast mode) states of a probe wave, respectively. (b) Schematic of the experiment. A pump pulse generates free carriers in the silicon by TPA and, consequently, induces a change of refractive index which propagates with the velocity of the pump pulse. The region with the red color gradient corresponds to the rising edge of the front. The orange arrow indicates the velocity of the pump pulse, while red, green, and blue arrows indicate the velocities of a wave packet of the probe at different times, respectively.

After injection of the pump pulse into the PhC waveguide, the pulse modifies the optical properties of the waveguide by generating FCs via TPA. In the switched zone, the refractive index of silicon, $n_0$, reduces by a quantity $\Delta n_{FC}$ that is proportional to the FC density $N_{FC}$, which in turn changes and blue shifts the dispersion curve. Therefore, a refractive index front constitutes a moving boundary between two photonic crystal zones with slightly different band diagrams[22]. When a probe wave packet is launched into the waveguide, the changes of its wave vector and frequency which are induced by the interaction with the co-propagating front are determined by the dispersion curve of the system, the propagation velocity of the front and the initial position of the probe wave vector and frequency in the band[22].

Here, we are interested in the particular situation where the probe wave packet ahead of the front cannot find states on the band of the switched PhC behind the front. Thus, the state of the probe wave, after interacting with the moving front, must remain in the initial band which means that an *intraband* transition takes place. This intraband transition manifests itself as a forward reflection from the front. The basic concept to induce this transition is schematically shown in Fig. 1(a). The phase continuity line defines possible states which satisfy a continuous phase at the refractive index front. Thus the moving front can only excite states that lie on the phase continuity line. The ratio of the possible frequency shift $\Delta\omega$ and the wave vector shift $\Delta k$ is equal to the front velocity. This relation can be derived using the phase conservation under Lorentz transformation [22], phase evolution integrals [26] or from the Doppler equation [27]. Subsequently, by utilizing the flexibility in dispersion designs, a variety of indirect transitions can be envisaged by changing the intersection points of phase continuity line with the bands of the photonic crystal before and after the front. We are interested in the particular situation when the phase continuity line does not cut through the band of the perturbed photonic crystal and thus there is no states behind the front that can be excited.

The intraband transition can be achieved by setting the pump pulses at the knee of the solid band at the frequency where group velocity corresponds to the slope of the phase continuity line, as illustrated in Fig. 1. The absence of states in the perturbed waveguide force the probe wave to stay in the unswitched zone of the PhC and is thus reflected at the moving front. This reflection is accompanied by a large frequency shift which, most importantly, can be achieved with a very small band shift. The final state of the probe ($\omega_2$, $k_2$) is determined graphically from the crossing point of the phase continuity line and the solid band. Fig. 1(b) shows a schematic illustration of this process. The fascinating fact is that a probe wave packet initially propagating slower than the approaching pump pulse, upon interaction with the moving front, which is dragged by the faster pump pulse, is accelerated and finally escapes from the moving front in the forward direction.

However, implementing the configuration shown in Fig. 1(a) by choosing the pump pulse to lie at the knee of the dispersion band has some experimental drawbacks. Firstly, due to the high intensity of the pump pulse and its center frequency close to that of the probe wave, it is difficult to detect the shifted probe after interaction. Secondly, it is also challenging to distinguish the intraband transition from third order nonlinear processes such as four wave mixing (FWM) which would cause spectrally similar signals. Thus, the pump should be positioned at some other frequency where the group velocity is the same as the slope of the phase continuity line.

Normally, slow light waveguides are optimized to obtain large bandwidth of constant small group velocity [20,21]. But slow light PhC waveguides can also be engineered to obtain a dispersion relation with equal group velocities at three different frequencies [21]. This engineered waveguide can be used to excite pump pulses with the required group velocity at a frequency distant from the initial and final frequencies of the signal.

Thus to implement intraband transitions, we fabricated a single line defect PhC waveguide consisting of a hexagonal lattice of air holes in silicon. The silicon PhC waveguide has a length of 396 µm and is connected to the edges of the chip by inverse tapers and polymer waveguides. The detailed design and measurement method of these waveguides were given in Refs.[21,22], [see also Supplemental Material]. The measured linear transmission (including coupling loss) and the group index of the TE-mode of the waveguide are shown in Figs. 2(a) and 2(b), respectively. Orange and red points in Fig. 2(b) indicate the wavelengths and group indices of pump pulse and probe wave, respectively. As clearly seen from the spectrum, the pump pulse at ≈1539 nm has a velocity matched wavelength at ≈1529.5 nm ($n_g^f = 30$). If we set the probe wave to 1531.5 nm, which is a bit longer than this matched wavelength, it will propagate with a slightly smaller group velocity than the refractive index front moving with the pump pulse. We will then expect an intraband transition to 1528.5 nm (Fig. 2, blue dot), which is the shortest of the three interacting wavelengths. The shifted probe wave at 1528.5 nm employs the smallest group index and therefore is going to escape from the approaching refractive index front, while staying in the unswitched zone, thus in the unswitched band. This arrangement is different to the scenario which features the pump pulse wavelength or frequency in between the respective values for the probe and shifted probe waves -easy to confuse with a FWM process. The modified scenario is now based on the pump pulse wavelength being the longest one and the shifted probe wavelength being the shortest one. This choice of parameters clearly rules out a FWM-process which, otherwise, would have competed with our novel intraband transition. Figure 2(c) shows the calculated dispersion band corrected by the slope corresponding to the pump pulse group velocity. Due to very small band diagram shift the curves cannot be presented in the original form similar to Fig. 1a as the lines would lie very close to each other. The slope corrected wavenumber k′ at a frequency ω is calculated as $k'(\omega) = k(\omega) - (\omega - \omega_{pump})/v_f - k(\omega_{pump})$. The orange line represents the phase continuity line with a slope equal to the traveling velocity of the pump pulse[22], which leads to a horizontal line in the corrected representation. By choosing the group index of the probe wave to be $n_g^s = 33$, we expect an intraband transition with a blue shift of ≈ 3.1 nm.

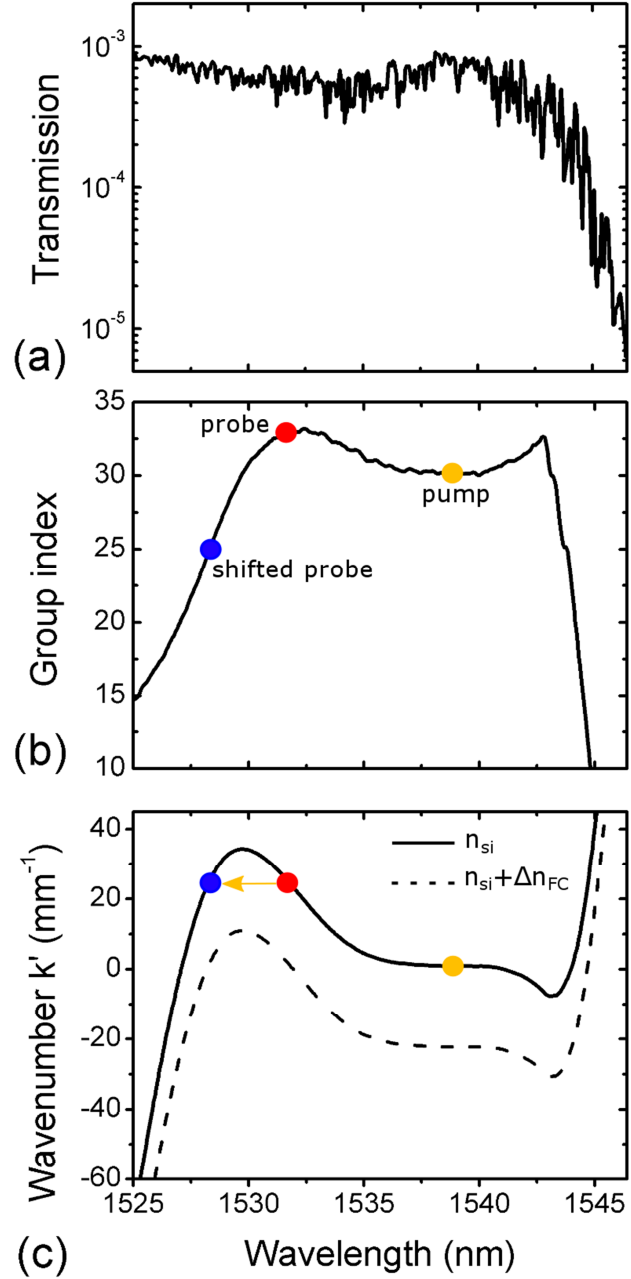

Figure 2: (a) Linear Transmission and (b) Measured group index of the engineered and fabricated slow light silicon PhC waveguide. (c) The calculated dispersion band from the measured group index in (b). The wave vector corrected by the slope of the pump group velocity is presented. Due to very small band diagram shift the curves cannot be presented in the original form similar to Fig. 1a. Solid curve represents the original dispersion band of waveguide mode with refractive index $n_{si}$, while the dashed curve indicates the waveguide mode with refractive index change of $-8 \cdot 10^{-4}$. Dots indicate the locations of the input wavelengths of probe wave (red dot), pump pulse (orange dot), and the expected output wavelength of the shifted probe wave (blue dot) after the intraband transition took place. The orange line represents the phase continuity line with a slope equal to the traveling velocity of the pump pulse. Due to the slope correction, this phase continuity line now appears horizontal.

**Experiment**

The optical measurement proceeds by launching 6 ps-long pump pulses derived from 100 MHz repetition rate mode locked fiber laser into a 396 µm-long slow light silicon PhC waveguide at a center wavelength of 1539 nm with a group index of $n_g^f = 30$. We also feed in the low power probe as a continuous wave (CW) of light, which co-propagates with the index front in the waveguide with a slightly slower group velocity. More details about the experimental setup can be found in the supplemental material and in Ref. [30]. The occurrence of an intraband indirect transition in our waveguide can be verified by recording the optical spectra of the output probe light with and without pump pulses present and as function of the pump pulse power.

**Results**

Figure 3 shows the experimental output spectra recorded for an on-chip CW probe wave power of ≈ 6 µW with and without the pump pulses. Black traces correspond to the output spectra of the CW probe that has not interacted with the pump pulses, blue traces relate to the pump pulses alone, and other colors refer to the cases when both the pump pulses and the CW probe wave were present (the curves are shifted for clarity, the noise level is -60 dBm). Figure 3(a) demonstrates the measured output spectra for the blue shifted probe wave at a wavelength of 1531.5 nm (group index of $n_g^s = 33$) as a function of the pump pulse peak power. With the pump pulses present, spectral components clearly appear which are blue-shifted with respect to the initial probe wavelength. Wave packets of the CW probe were approached by the index front and were transmitted through it or reflected from it in forward direction.

The group velocity of the pump pulses is fixed, and, accordingly, the slope of the phase continuity line is defined as well. Therefore, the induced transition of the probe wave is governed by its initial wavelength location on the band diagram and by the magnitude of the band shift. We estimate that the pump power at the input of the slow light waveguide is reduced by coupling losses of approximately -4dB[22]. When the pump pulse peak power in the waveguide is low (≈ 2 W), i.e. when Δn is small, we can only notice one blue-shifted peak at ≈1530 nm (marked by black arrow in Fig. 3(a)), which is attributed to the interband transitions into the slightly shifted band into which the phase continuity line cuts. This phenomenon was already discussed in Ref. [22]. However, when we increase the pump pulse power, the band shift also increases and we can observe the appearance of another peak with a much larger blue shift. At sufficient Δn, the phase continuity line can reach the shorter wavelength position in the same band without cutting into the shifted band. This intraband transition (marked by blue arrow) is realized at the estimated pump pulse peak power of 6.2 W and pulse duration of 6 ps which leads theoretically to FC generation of $5\cdot10^{17}$/cm³ and a refractive index change of approximately $-(1-2)\cdot10^{-3}$ at the input of the waveguide (see supplementary materials). The refractive index change can be also estimated from the wavelength position of the dispersion wave generated by the pump alone, which is observed as a peak at 1531.5 nm in the curves when only the pump is entering the waveguides. This estimation is also conducted in the supplementary materials and leads to the estimated refractive index change of $0.8\cdot10^{-3}$. We will use this estimation of the refractive index throughout the manuscript.

With increasing power the signal height caused by those probe wave packets subject to *intraband* transition increases, the signal height of those wave packets which underwent the *interband* transitions decreases. As the pump pulse power further increases, the wavelength shift saturates as additional pump power only further separates the bands but has no effect on the starting and end points in the band diagram which indicate the intraband transition. The center wavelength of the maximum blue-shifted peak is evaluated to be at ≈ 1528.5 nm. We calculate the center wavelength of the shifted probe spectrum using eq. (3) in Ref. [22]. The measured wavelength shift of −3 nm matches very well the value of −3.1 nm we expect from the calculated dispersion band in Fig. 2(c). In the time domain, a probe wave packet which is travelling slower than the index front is converted to a wave packet travelling at a new frequency. The shifted probe then travels faster in the unswitched zone of the waveguide than the approaching front (see Figs. 2b and 2c) and thus escapes in forward direction. The intraband transition presented here is not reversible. A probe wave launched at 1528.5 nm from behind the pump pulse travels faster than the pump pulse, undergoes an interband transition and turns out to be blue shifted again as it penetrates the refractive index front and dives into the unswitched zone. These results are presented in the supplementary materials.

Figure 3(b) displays the recorded output spectra of the blue-shifted probe wave for three different input probe wavelengths at 1530.8, 1531.5, and 1532 nm. The group velocities of these input probe wave packets are always slower than the group velocity of the pump pulse, thus than the velocity of the index front. As expected from the high steepness of the band shown in Fig. 2(c), the maximal wavelength of the shifted probe wave is not significantly affected as the input probe was slightly tuned towards longer wavelengths. A maximum blue-shift of ≈3.4 nm is achieved for the input probe at wavelength of 1532 nm. This is a more than three times larger wavelength shift as observed with indirect interband transitions with 50% less FCs[22], and more than ten times larger than what is expected from the instantaneous refractive index change of only $0.8 \cdot 10^{-3}$. This indicates the vast potential of intraband photonic transitions for light manipulation on chip.

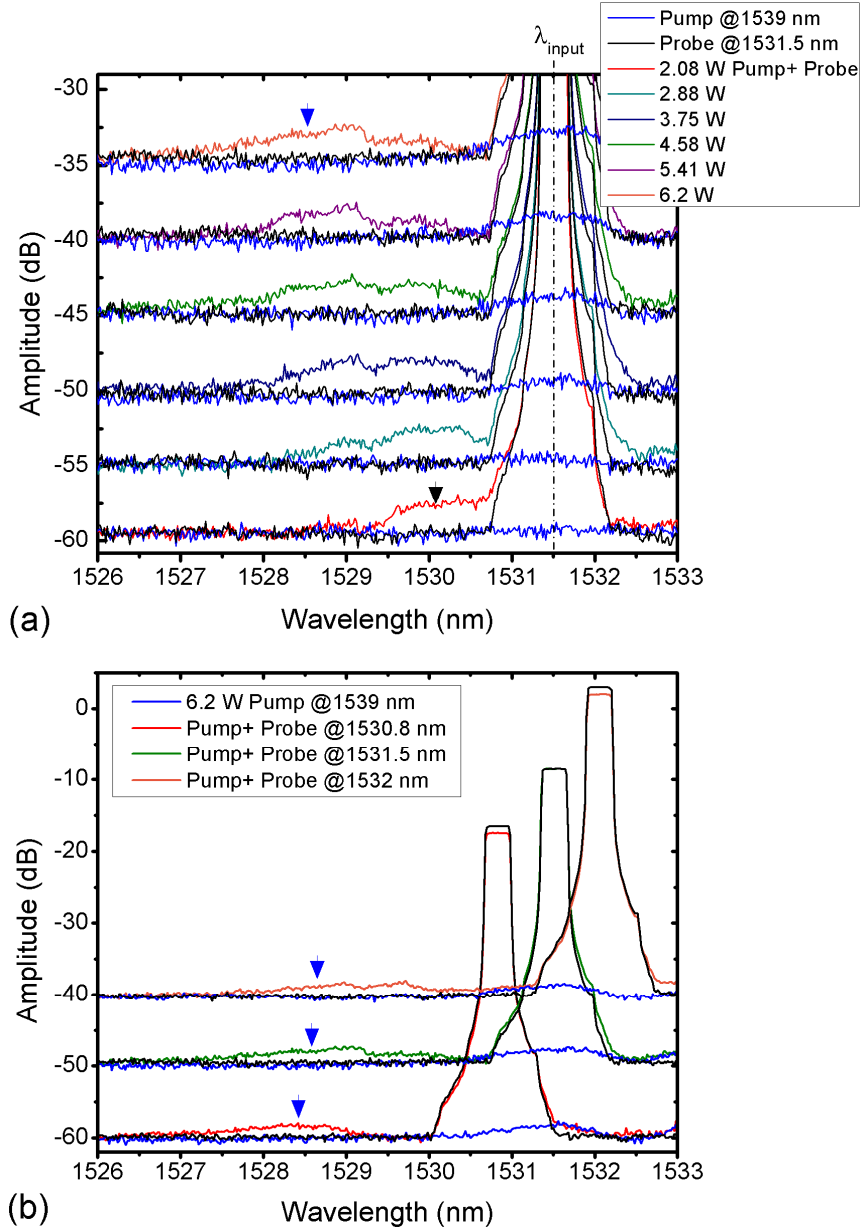

Figure 3: Experimental spectra recorded at the output of our 396 µm-long slow light silicon PhC waveguide. (a) Pump pulse peak power-dependent output spectra for the blue shifted light of an input probe wave at a wavelength of 1531.5 nm. Shown is the spectrum of the CW probe that has not interacted with the pump pulses (black traces), the spectrum of the pump pulses alone (blue traces), and shown are the spectra of both the pump pulses and the CW probe (other colors). Black and blue arrows illustrate the interband and intraband photonic transitions, respectively. (b) Spectra for different input probe wavelengths. The different curves in (a) and (b) are shifted by 5 and 10dB, respectively for clarity, the noise level corresponds to -60dBm.

## Discussion

Here we discuss the results obtained in Fig. 3 and give an explanation for the observed blue shifted probe wave signal. Figure 4(a) shows a schematic representation of the dispersion band of the PhC waveguide. Red and orange dots again indicate the locations of the input probe wave and the pump pulses, respectively. The spatiotemporal change in Δn is schematically demonstrated in Fig. 4(b) (left). The red filling color indicates the z-coordinate range of the waveguide where the refractive index has been switched already (switched zone). As the time evolves to the left the switched zone dimension extends to positive z-coordinate values. Eventually, the waveguide is completely switched. As the FC lifetime in silicon photonic crystal waveguides is in the order of 100 ps[31,32], there is no recombination taking place in the time frame of the front propagation. The graded reddish area indicates the fact that the time function of the pump pulse power employs a finite steepness, thus the refractive index front is gradual in any position as the pump pulse transforms the waveguide medium. The two horizontal dotted lines represent the spatial width of the index front. The velocity of the refractive index front, indicated in the diagrams (b), (d) and (f) as the slope of any line of equal reddish color tone in the graded part, is the same as the group velocity of the pump pulse and thus is equal to the slope of the band at pump frequency and slope of the phase continuity line. The induced transitions of the input probe wave into the photonic bands (Figs. 4(c), (e)) and the corresponding trajectories of its wave packets (Figs. 4(d), (f)) for two different maximal Δn values are illustrated, respectively.

To explain the obtained results, we choose to view CW probe wave as being composed of individual wave packets located at different z-coordinate positions at a given point in time. These wave packets then interact with the front differently, depending on their relative positions to the front at a given time. Their trajectories are shown in Fig. 4 (d) and (f) by solid and dotted black lines. Before the faster approaching index front encounters the wave packet, this wave packet moves along a straight line with constant group velocity of the probe. Upon contact with the front, the probe wave packet first penetrates into the zone of reduced refractive index and thus experiences an indirect transition to the new frequency and wave vector. This leads to change of the group velocity which determines how the wave packet further propagates. If the maximal refractive index change of the front is sufficient to achieve an intraband transition to a band diagram position of larger group velocity, then the probe wave packet penetrates the front up to a position where its group velocity has gradually increased to match that of the front. After that, the probe wave packet starts to recede from the front, again, as its group velocity further increases and it escapes in the forward direction. A useful approximation of the wave packet trajectory inside the front can be derived for the case where the band of the unswitched waveguide upon switching is just vertically shifted in frequency proportionally to the refractive index change and, assuming further, that the front has a graded refractive index, linearly growing along the negative z-direction. In this case, it can be shown that the wave packet trajectory in time and space has the same shape as the band in wave vector and frequency coordinates. This correspondence comes from the equivalency of the slopes in both curves as $d\omega_{band}/dk_{band}$ and $dx/dt$ are both the group velocity of the wave packet. Also the frequency change of the wave packet along the trajectory is directly proportional to the time spent inside the front $\Delta t$:

$$\Delta \omega = v_f \Delta k = \Delta \omega_{shift} \cdot \Delta t/\tau \tag{1}$$

where $v_f$ is the front velocity, a.k.a. the group velocity of the pump pulse. $\Delta k$ is wave vector change, $\Delta \omega_{shift}$ is the maximal shift of the band diagram produced by the front and $\tau$ is the rising time of the front. In our experiments the pump pulse duration and thus the rising time of the front is fixed, but the maximal band shift increases as the pump pulse power increases.

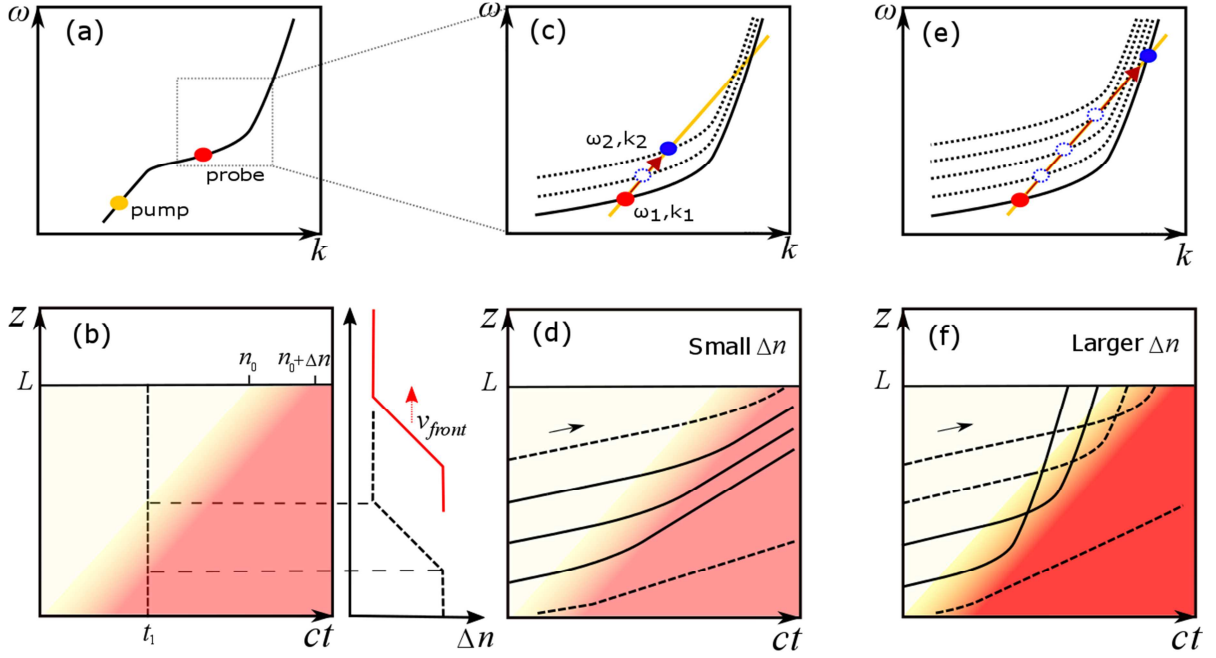

Figure 4: (a) Schematic representation of the dispersion band of our PhC waveguide. Red and orange dots indicate the locations of the input probe wave packets and the pump pulse, respectively. (b) Spatiotemporal change of Δn. The red filling color indicates the z-coordinate range of the waveguide where the refractive index has been switched already (switched zone). The graded reddish area indicates the fact that the time function of the pump pulse power employs a finite steepness, thus the refractive index front is gradual in any position as the pump pulse transforms the waveguide medium. The two horizontal dotted lines represent the raising length of the index front. The velocity of the refractive index front, indicated in the diagrams (b), (d) and (f) as the slope of any line of equal reddish color tone in the graded part, is the same as the group velocity of the pump pulse thus equal to the slope of the band at pump frequency. (c) and (e) show the dispersion bands and the corresponding interband and intraband photonic transitions of the input probe caused by insufficient (dashed blue dots) and by sufficient (solid blue dots) values of Δn, respectively. (d) and (f) Schematic illustrations of the probe wave signal trajectories for the induced transitions in (c) and (d), respectively. Black lines in (d) and (f) represent wave packets of the CW input probe. Solid lines illustrate the wave packets which undergo complete interband (d) and intraband (f) transitions, while the dashed lines depict the wave packets which only undergo incomplete transitions.

First, we consider the situation when $\Delta\omega_{shift}$ is small i.e. at low pump pulse power (red curve in Fig. 3(a)). In this case, the initial band only shifts slightly to a higher frequency and a probe wave packet with an initial state ($\omega_1$, $k_1$), upon interaction with front, moves along the phase continuity line to a final state ($\omega_2$, $k_2$) in the shifted band (Fig. 4(c)). This is the interband transition case. Then, the expected wavelength change is relatively small for the given interaction time, such is the corresponding group velocity change inside the front. The group velocity change in this case is insufficient in order to transform the probe packets and to accelerate them away from the front. Thus, the wave packets are transmitted through the front into the switched zone where they further propagate in the forward direction behind the front, at a slightly higher frequency and at a slightly higher group velocity as compared to before. After the injection of the front, some of these probe wave packets will undergo complete transitions (solid lines) as they experience maximal possible time in the front. However, there are other probe wave packets which undergo incomplete transitions (dotted lines). The incomplete transitions are attributed to the fact that the packets are either already inside the refractive index front at the input (lower dotted line), or they exit the slow light waveguide before complete transition could take place (upper dotted line).

Next, the situation where the intraband transition occurs and the probe wave packets completely escape from the front is considered. The band diagram shown in Fig. 4(e) indicates that the intraband transition can be realized easily when $\Delta\omega_{shift}$ is large (the band shift is large). This behavior suitably explains the occurrence of the intraband transition indicated by the blue arrow in Fig. 3(a) as we increase the power of the pump pulses. The trajectories of the probe wave packets for this case are illustrated in Fig. 4(f).

In the presented description we did not take into account the decay of the pump power in the slow light waveguide due to linear or nonlinear absorption and due to scattering. Such losses diminish the FC concentration thus lead to the reduction of $\Delta\omega_{shift}$ along the waveguide. Therefore, such losses reduce the total effect of intraband transitions along the full waveguide length.

We also observe a peak generated only by the pump pulses at approximately 1531.5 nm. This peak can be attributed to the special case of the dispersion wave (DW)[28,29,33,34]. We explain the appearance of this DW by self-induced indirect transition of the pump pulses (see supplementary materials).

As we use a CW probe, only a small fraction of the total probe light can be converted within the finite length of PhC waveguide. The fraction η of the CW probe wave power that enters the front and is frequency converted is:

$$\eta = \nu_{rep} \cdot \left[\left(\frac{n_g^s - n_g^f}{c}\right) \cdot L + \tau\right] \quad (2)$$

Where $\nu_{rep}$ is the repetition rate of the pulses, $c$ is the velocity of light in vacuum, $L$ is the length of the PhC waveguide and $\tau$ is the rising time of the front. The first part in the bracket corresponds to the part of CW probe that entered the front within the waveguide length $L$ and the second term relates to the CW probe wave that entered the waveguide when the front has just partly evolved for a fraction of its rise time $\tau$. We approximate the rise time of the front $\tau = 6ps$ approximately by the duration of the pump pulse. For $\nu_{rep} = 100\ MHz$, $n_g^f = 30$, $n_g^s = 33$ and $L = 396\ \mu m$, the maximally transformable fraction of the incoming CW probe power due to the interaction with single pump pulses arriving at 100 MHz repetition rate is $1\cdot 10^{-3}$ and this value corresponds to a 100% reflection efficiency. From the measurements shown in Fig. 3(a) a pulsed pump with peak power of 6.2 W acts on the on-chip CW probe with power of ≈ 6 µW and causes a forward reflected and frequency transformed signal pulses with average power of ≈ 2nW. This transformed power corresponds to a fraction of the transmitted CW probe power of ≈ $3.5\cdot 10^{-4}$. The ratio of this transformed probe power fraction to the maximally transformable probe power fraction of $1\cdot 10^{-3}$ equals 35% and marks the efficiency of the intraband transition caused by reflection at the relativistically propagating plasma mirrors. The missing 65% can be attributed to the absorption of the probe wave packets by FCs during their interaction with the plasma mirror front. This highly effective reflection employing only small pump power levels and thus small refractive index changes becomes possible due to the novel indirect intraband optical transition in our dispersion engineered slow light waveguide.

We would like to discuss here briefly the advantages of the frequency shift induced by the FC front in comparison to idler generation in the FWM configuration. First, the time span of the signal wave shifted to new frequency is not limited by pulse duration as in FWM but depends on both the group velocity differences between the pump and the probe and on the interaction length. Thus, with such a FC front we can envisage to shift large portions of signal, for example a packet of binary optical signal information. For example, if we assume a 1mm long slow light waveguide and a group index difference between the pump and the signal of 30, then the switching time will be 100 ps. This value is much larger than what can normally be achieved in a FWM process, which is limited by the picosecond, or shorter, pump pulse durations. In addition, in our approach, the pump can be positioned outside of the signal frequency window. Whereas in the FWM approach the pump frequency always has to lie between signal frequency and shifted frequency thus causes a bigger cross-talk problem. And the third advantage is the fact that to change the frequency shift neither power nor frequency of the pump have to be adjusted. The different frequency shifts can be achieved with the same pump in waveguides with different dispersion. These advantages make the interband optical transitions a viable option for all optical signal routing in wavelength-division-multiplexing networks[35].

The free carrier plasma mirror is asymmetric in its reflection properties. A fast probe wave packet that would approach the front from the rear side would be transmitted through the front and dive into the unswitched zone as it is blue shifted, too (see supplementary material).

## Conclusion

In conclusion, we have experimentally demonstrated a 35% optical-wave reflection and -3.4 nm wavelength shift by interaction with a relativistic FC plasma mirror generated, confined and propagating inside a 400 µm long dispersion engineered silicon slow light PhC waveguide. The moving mirror was generated by two photon absorption of 6 ps long pump pulse with a peak power of 6.2 W. The forward reflection of the probe wave packets co-propagating with the plasma mirror are accompanied by a frequency upshift. The presented reflection becomes possible due to a novel indirect intraband optical transition in a slow light waveguide. Under these special conditions the incident light does not find states beyond the front and has to reflect from it. Our novel concept bears various technical advantages such as that no vacuum conditions and no ultrahigh power lasers are required. Most importantly, it allows efficient reflection at small FCs concentration. In contrast to the previously demonstrated plasma mirrors produced by extremely high pump power levels, the effect presented here is based on the fascinating dispersion properties of carefully engineered silicon slow light PhC waveguides which allow to control the speed of the plasma mirror with respect to the group velocity of the probe wave packets. We show also that the FC plasma mirror is asymmetric in its reflection properties. It will accelerate and reflect a slow probe wave packets propagating in front of it and will transmit a fast probe wave packets initially behind it. Due to the fact that the pump, probe and shifted probe are all at 1.5 µm wavelength, the presented effect allows not only fundamental studies of relativistic effects but also open new possibilities for frequency manipulation and all optical switching in optical telecommunication.

## Acknowledgements

M.A.G, D.J., A.Yu.P. and M.E. acknowledge the support of the German Research Foundation under Grant No. EI 391/13-2, and appreciate the support of CST, Darmstadt, Germany, with their Microwave Studio Software. M.A.G, D.J., A.Yu.P. and M.E. acknowledge the support of Michel Castellanos Muñoz in preparing the grant proposal. J.L acknowledge the supports of the Ministry of Science and Technology of China (2016YFA0301300) and National Natural Science Foundation of China (11674402)

## Author contributions

M.A.G performed the simulations and measurements, J.L and L.F fabricated the PhC waveguide, T.F.K., A.Yu.P. and M.E. supervised the project, M.A.G, A.Yu.P., D.J., T.F.K. and M.E. analysed the results and wrote the paper.

## Competing financial interests

The authors declare no competing financial interests.

# Supplemental material of "Highly effective relativistic free carrier plasma mirror confined within a silicon slow light photonic crystal waveguide"

Mahmoud A. Gaafar,[1,2,*] Dirk Jalas,[1] Liam O'Faolain,[3,4,5] Juntao Li,[6,+] Thomas F. Krauss,[7] Alexander Yu. Petrov[1,8], and Manfred Eich[1,9]

[1]*Institute of Optical and Electronic Materials, Hamburg University of Technology, Hamburg 21073, Germany*
[2]*Department of Physics, Faculty of Science, Menoufia University, Menoufia, Egypt*
[3]*SUPA, School of Physics and Astronomy, University of St. Andrews, St. Andrews, Fife KY16 9SS, United Kingdom*
[4]*Tyndall National Institute, Lee Maltings Complex, Dyke Parade, Cork, Ireland*
[5]*Centre for Advanced Photonics and Process Analysis, Cork Institute of Technology, Cork, Ireland*
[6]*State Key Laboratory of Optoelectronic Materials & Technology, Sun Yat-sen University, Guangzhou 510275, China*
[7]*Department of Physics, University of York, York, YO105DD, United Kingdom.*
[8]*ITMO University, 49 Kronverkskii Ave., 197101, St. Petersburg, Russia*
[9]*Institute of Materials Research, Helmholtz-Zentrum Geesthacht, Max-Planck-Strasse 1, Geesthacht, D-21502, Germany*
*\*Corresponding author: mahmoud.gaafar@tuhh.de*
*+Corresponding author for slow light sample fabrication: lijt3@mail.sysu.edu.cn*

**Structure of the slow light silicon PhC waveguide and measurement setup**

The employed single line defect PhC waveguide has been fabricated on a Silicon-on-Insulator substrate with slab height of 220 nm. The length of the waveguide is 396 µm and is connected to the edges of the chip by inverted tapers with polymer waveguides to reduce the coupling losses. In this study, the lattice constant of the PhC is 404 nm and the air-holes diameter is 240 nm. In order to achieve a waveguide with required dispersion, the first row of holes directly adjacent to the waveguide has been shifted 50 nm away from the waveguide center[1]. The oxide underneath the waveguide has been removed, in order to achieve a symmetric air cladding[1,2]. Coupling into the structure is carried out in an end fire configuration, with coupling losses of approximately -4dB per facet.

The experiment is arranged as displayed in Fig. S1. A mode-locked fiber laser (Menlo system) with a 100 MHz repetition rate delivers pulses of approx. 100 fs duration with a wide spectrum ranging from 1500nm–1600nm. The output is connected to a grating filter with adjustable bandwidth and center frequency followed by an erbium-doped fiber amplifier (EDFA). Pulses with tunable center wavelength between 1525 nm –1590 nm and bandwidth between 0.1 nm – 4 nm (33 ps – 2 ps) can be produced from this configuration. The probe light is derived from a tunable diode laser (Photonetics PRI). It delivers up to 8 mW of CW light, tunable from 1500 to 1600 nm. The adjustment of optical power and polarization is carried out with a combination of half-wave plates and polarizers. The pump pulse is subsequently combined with the probe light through a 50/50 beam splitter. The optical spectra are measured by the optical spectrum analyzer.

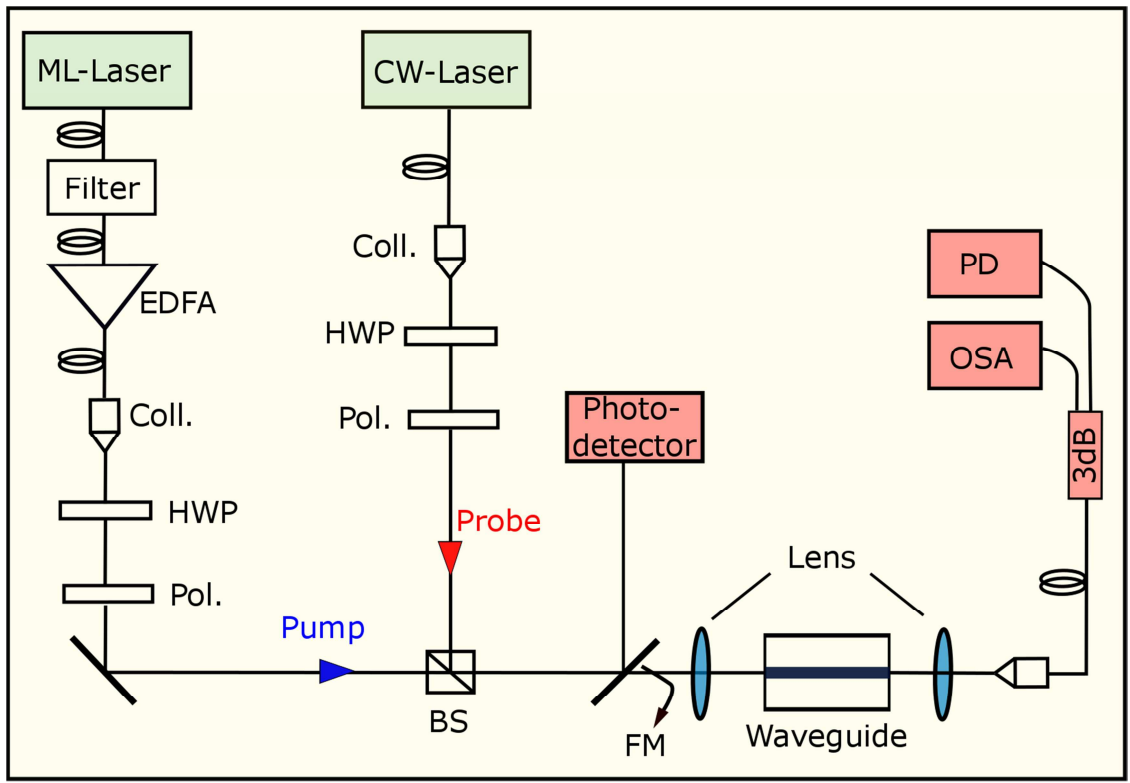

Figure S1: Schematic of the experimental setup. EDFA, erbium-doped fiber amplifier; Coll., collimator; Pol., polarizer; HWP, half-wave plate; BS, beam splitter; FM, Flipping mirror; OSA, optical spectrum analyzer; and PD, photodiode.

**Free carrier generation and index change of an optical mode**

The time dependence of the free carrier (FC) density $N(t)$ at a fixed point in space generated by two-photon absorption (TPA) of a pump pulse propagating in a slow light silicon waveguide is described by[3]:

$$\frac{dN(t)}{dt} = \frac{\beta}{2hf}\left(\frac{n_g}{n_{si}}\right)^2 I^2(t) - \frac{N(t)}{\tau_{FC}} \qquad (1)$$

The first term on the right-hand-side of this equation describes the generation of free carriers via the TPA process, while the second term describe the decay of the generated free carriers, characterized by the lifetime $\tau_{FC}$. Here, $h$ is Planck's constant, $\beta = 0.7 \: cm/GW$ is the TPA coefficient in silicon at a frequency $f = 193.4 \: THz^4$, $n_g$ is the pump group index, $n_{si}$ is the silicon refractive index, and $I(t)$ is the pump pulse intensity.

The FC induced refractive index change can be described by the empirical relation presented by Soref et al.[5]:

$$\Delta n(t) = -8.8 \cdot 10^{-22} N(t) - 8.5 \cdot 10^{-18} N(t)^{0.8} \qquad (2)$$

We solve equation (1) numerically assuming a Gaussian pulse intensity $I(t) = I_0 \exp(-4\ln 2 \, t^2/t_{pulse}^2)$ with FC life time $\tau_{FC} = 1 \: ns$. The maximum of the generated FC density is given by:

$$N_{FC,max} = \frac{\beta}{2hf}\left(\frac{n_g}{n_{si}}\right)^2 I_0^2 \sqrt{\pi} \frac{t_{pulse}}{\sqrt{8\ln 2}} \qquad (3)$$

When a small perturbation of the refractive index is applied, the frequency of an eigenmode will change according to[6]:

$$\Delta\omega = -\frac{\omega}{2} \frac{\int \Delta\varepsilon(\vec{r})|\vec{E}(\vec{r})|^2 d\vec{r}}{\int \varepsilon(\vec{r})|\vec{E}(\vec{r})|^2 d\vec{r}} \qquad (4)$$

For homogeneous spatial distributions for $\Delta\varepsilon$ and $\varepsilon$ one gets from equation (4) the relation:

$$\frac{\Delta\omega}{\omega} = -\frac{\Delta n}{n} \qquad (5)$$

Considering the dynamic switching configuration in our PhC waveguide, special attention should be paid to carrier diffusion in silicon. The spatial distribution of $\Delta n$ is restricted to the center region of the waveguide up to the surrounding first row of holes, which inhibits carrier diffusion out of the center region of the waveguide[7]. Assuming a small $\Delta n$ which is constant over the center region and zero elsewhere, compared to a spatially homogenous change of the refractive index, this assumption will not lead to a considerable change of the slope of the shifted band diagram, and one gets from equation (4)[6]:

$$\frac{\Delta \omega}{\omega} = -\frac{\Delta n}{n} \cdot \text{(fraction of } \int \varepsilon |E|^2 \text{ in the perturbed regions)} \tag{6}$$

From this equation we can estimate the expected frequency shift of the optical mode according to the maximum refractive index change $\Delta n$. For the presented slow light waveguide the fraction is close to one.

For $t_{pulse} = 6\ ps$, $n_g = 30$, a pulse peak power of 6.2 W, and by taking into account an effective mode area of the pump pulse of $0.5\ \mu m^2$, we estimate the maximum index change at the input of the waveguide of $\approx -(1\text{-}2)\cdot 10^{-3}$ and frequency shift of $\Delta\omega \approx 90$ GHz, which corresponds to the maximum of the generated FC density of $\approx 5\cdot 10^{17}$/cm³. This value of refractive index change corresponds well with a refractive index change estimation based on the consideration of the generated dispersion wave (as we show later in Figs. S3 and S4). We should also mention, that the pump power decays in the slow light waveguide due to absorption and scattering, and accordingly does the $\Delta n$.

**The intraband transition is not reversible**

The FC plasma mirror is asymmetric in its reflection properties. Due to the long lifetime, the FC front does not cause an opposite slope of the refractive index front at its rear side. Thus, the fast probe wave packet approaching the front from the rear side is transmitted through the front and is blue shifted again. This is an interband transition, again, however, this time, the starting point originates from the switched band, then the probe wave packet is guided along the phase continuity line and finally lands at a higher frequency on the unswitched band. Figure S2 shows the experimental output spectra recorded with and without pump pulses of an input probe at a wavelength of 1528.5 nm. Remarkably, with the pump pulses present, the induced indirect transition again leads to a blue shift of the probe frequency.

In contrast to the case presented in Fig. 3(a) where the probe wave packet was travelling slower than the front, now the probe wave packet is travelling faster than the front. Therefore the probe wave already enters the waveguide in its switched zone, thus the initial state of the probe wave ($\omega_1$, $k_1$) already lies on the switched band corresponding to $n_{si} + \Delta n_{FC}$. The final state of the probe ($\omega_2$, $k_2$) is determined graphically from the crossing point of the phase continuity line and the solid band, as schematically illustrated at the inset of Fig. S2, which indicates that the final state of the probe lies on the band corresponding to $n_{si}$.

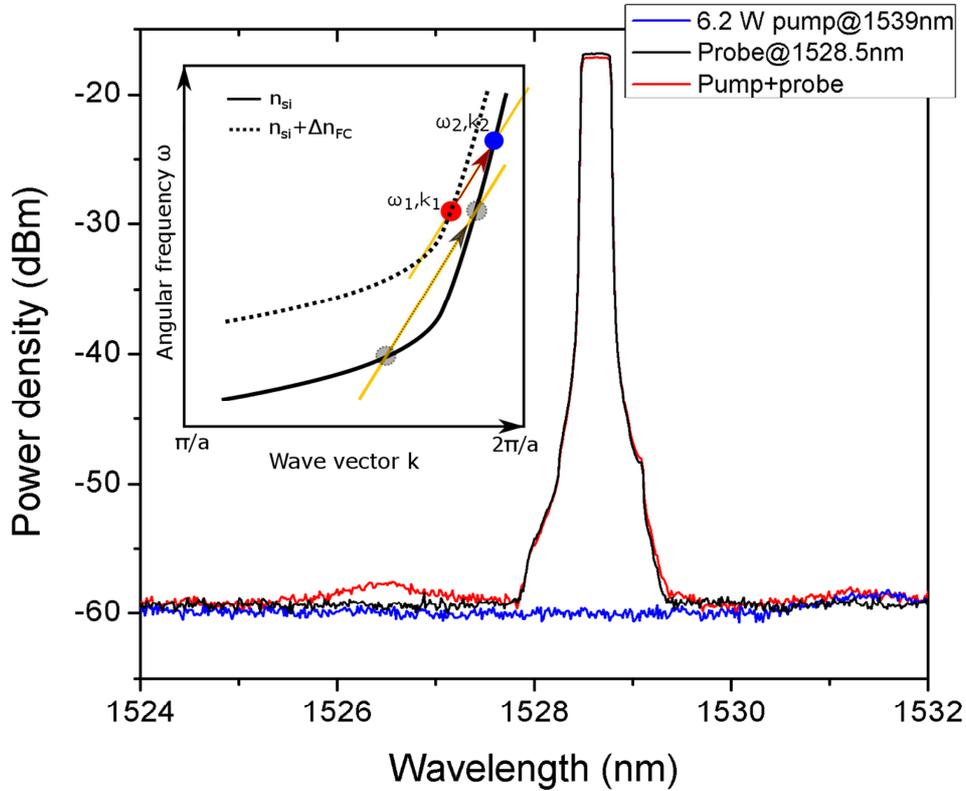

Figure S2. Experimental spectra recorded at the output of our slow light silicon PhC waveguide of an input probe at a wavelength of 1528.5 nm and pump peak power of 6.2 W. The spectrum of the CW

probe that has not interacted with the pump pulses (black trace), of the pump pulses alone (blue trace), and of both the pump pulses and the probe (red trace). Inset shows schematically the corresponding transition (red and blue dots). Grey dots represent the intraband transition when the faster pump pulse approaches the initially slower probe wave packet in the unswitched zone.

**Self-induced indirect transition of the pump pulses**

We also observed a peak solely generated by the pump pulses at approximately 1531.5 nm. This peak can be attributed to the special case of the dispersion wave (DW)[8–11]. Figure S3 demonstrates the measured pump pulse power-dependent output spectra for the blue shifted light of an input probe at a wavelength of 1531.5 nm and with a group index of $n_g^s = 33$. The 6 ps pump pulses are launched at a center wavelength of 1539 nm with a group index of $n_g^f = 30$. With increasing the pump pulses power, in addition to the blue-shifted probe, a self-induced blue-shifted peak of the pump pulses clearly appear at a wavelength of 1531.5 nm (blue traces and blue arrows). We explain the appearance of this DW by self-induced indirect interband transition of the pump pulses, as illustrated by the dotted arrow in Fig. S4. Appearance of DW at this wavelength, indicates that the maximum refractive index change we have is close to $8 \cdot 10^{-4}$.

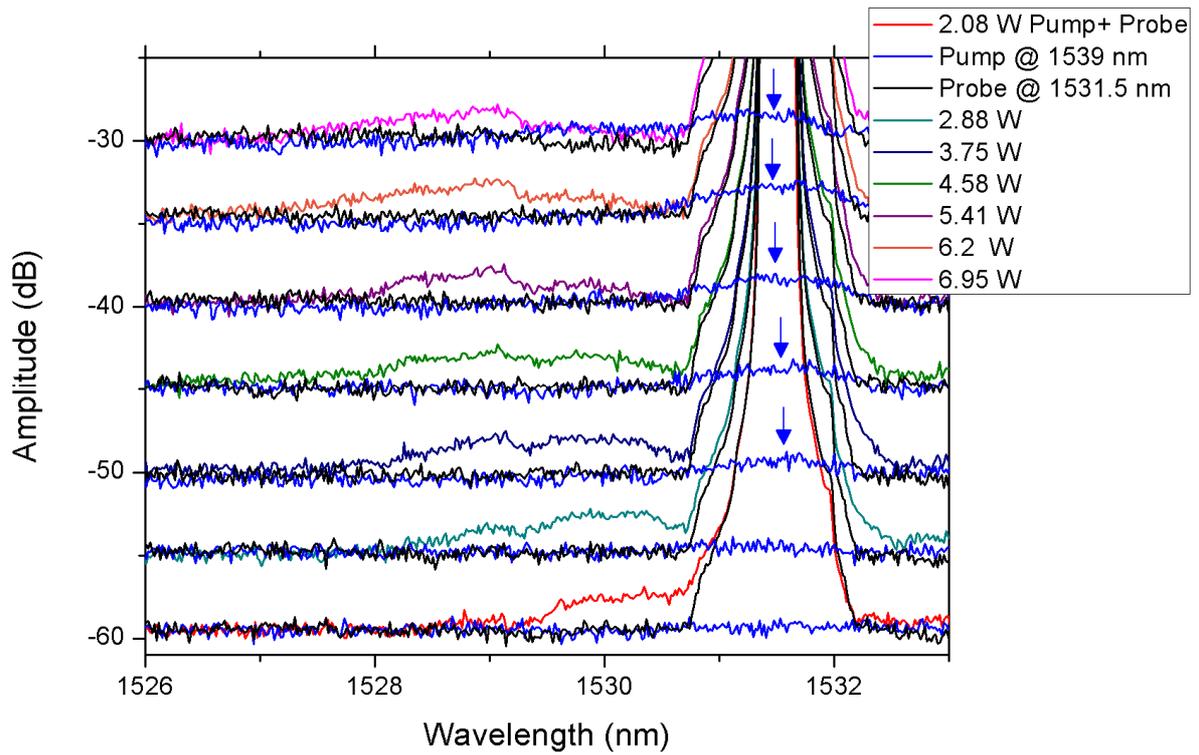

Figure S3: Pump pulse peak power-dependent output spectra for the blue shifted light of an input probe at a wavelength of 1531.5 nm. The spectra of CW probe that have not interacted with the pump pulses (black traces), of the pump pulses alone (blue traces), and of both the pump pulses and the probe (other colors). The different curves are shifted by 5 for clarity, noise level corresponding to -60 dBm.

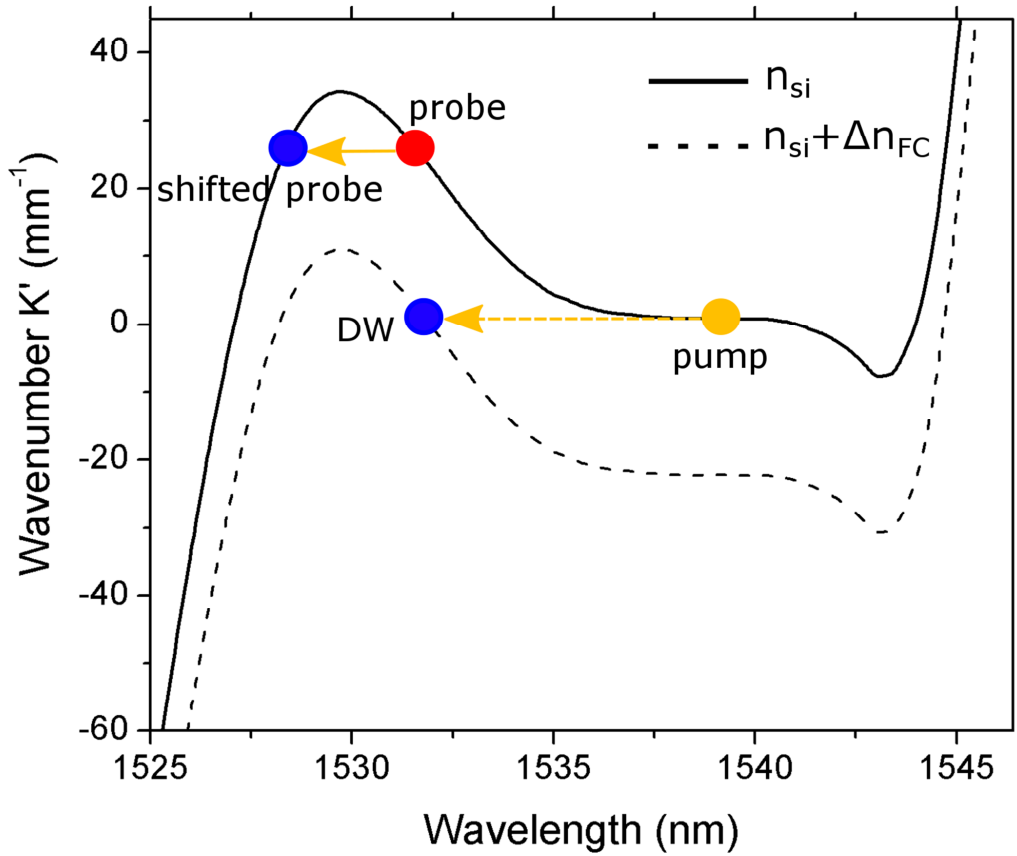

Fig. S4 The calculated dispersion band from the measured group index of our PhC waveguide corrected by the slope corresponding to the pump pulse group velocity. Solid curve represents the dispersion band of waveguide mode with refractive index $n_{si}$, while the dashed curve indicates the waveguide mode with refractive index change of $-8\cdot10^{-4}$. Color dots indicate the locations of the input wavelengths of probe (red dot), pump pulse (orange dot), and the expected wavelength of the shifted probe wave (blue dot). Orange arrows represent the phase continuity line with a slope equal to the traveling velocity of the pump pulse. Orange solid arrow demonstrates the intraband transition of the probe light, while the orange dashed arrow represents the self-induced indirect interband transition of the pump pulses.